\documentclass[prb,aps,preprint,nopacs,superscriptaddress,nofootinbib]{revtex4}
\usepackage{graphicx}
\usepackage{dcolumn}
\usepackage{bm}
\usepackage{bbm}
\usepackage{amsmath}
\usepackage{epsfig}
\usepackage{units}
\usepackage{esint}
\usepackage{braket}
\usepackage{color}
\usepackage{rotating}
\usepackage{float}

\def\ii{{\rm i}}  \def\ee{{\rm e}}
\def\rb{{\bf r}}      \def\vb{{\bf v}}
    \def\zz{\hat{\bf z}}
\def\kb{{\bf k}}    
    
\def\me{m_{\rm e}}  
    \def\Ab{{\bf A}}
  
\def\aC{\alpha_\text{C}}
\def\se{\sigma_\text{e}}  \def\sL{\sigma_\text{L}}
\def\te{\tau_\text{e}}  \def\tL{\tau_\text{L}}
\def\EE{{\mathcal E}}  \def\EEb{\vec{\mathcal E}}

\begin{document}

\title{From attosecond to zeptosecond coherent control of free-electron wave functions using semi-infinite light fields}
\author{G.~M.~Vanacore}
\thanks{Authors contributed equally}
\affiliation{Institute of Physics, Laboratory for Ultrafast Microscopy and Electron Scattering (LUMES), \'Ecole Polytechnique F\'ed\'eral de Lausanne, Station 6, CH-1015 Lausanne, Switzerland}
\author{I.~Madan}
\thanks{Authors contributed equally}
\affiliation{Institute of Physics, Laboratory for Ultrafast Microscopy and Electron Scattering (LUMES), \'Ecole Polytechnique F\'ed\'eral de Lausanne, Station 6, CH-1015 Lausanne, Switzerland}
\author{G.~Berruto}
\affiliation{Institute of Physics, Laboratory for Ultrafast Microscopy and Electron Scattering (LUMES), \'Ecole Polytechnique F\'ed\'eral de Lausanne, Station 6, CH-1015 Lausanne, Switzerland}
\author{K.~Wang}
\affiliation{Institute of Physics, Laboratory for Ultrafast Microscopy and Electron Scattering (LUMES), \'Ecole Polytechnique F\'ed\'eral de Lausanne, Station 6, CH-1015 Lausanne, Switzerland}
\affiliation{Technion Department of Physics}
\author{E.~Pomarico}
\affiliation{Institute of Physics, Laboratory for Ultrafast Microscopy and Electron Scattering (LUMES), \'Ecole Polytechnique F\'ed\'eral de Lausanne, Station 6, CH-1015 Lausanne, Switzerland}
\author{R.~J.~Lamb}
\affiliation{SUPA, School of Physics and Astronomy, University of Glasgow, Glasgow G12 8QQ, UK}
\author{D.~McGrouther}
\affiliation{SUPA, School of Physics and Astronomy, University of Glasgow, Glasgow G12 8QQ, UK}
\author{I.~Kaminer}
\affiliation{Institute of Physics, Laboratory for Ultrafast Microscopy and Electron Scattering (LUMES), \'Ecole Polytechnique F\'ed\'eral de Lausanne, Station 6, CH-1015 Lausanne, Switzerland}
\affiliation{Technion Department of Physics}
\author{B.~Barwick}
\affiliation{Ripon College, 300 W. Seward St., Ripon, WI 54971, United States}
\author{F.~Javier~Garc\'{\i}a~de~Abajo}
\email[E-mail: ]{javier.garciadeabajo@nanophotonics.es}
\affiliation{ICFO-Institut de Ciencies Fotoniques, The Barcelona Institute of Science and Technology, 08860 Castelldefels (Barcelona), Spain}
\affiliation{ICREA-Instituci\'o Catalana de Recerca i Estudis Avan\c{c}ats, Passeig Llu\'{\i}s Companys 23, 08010 Barcelona, Spain}
\author{F.~Carbone}
\email[E-mail: ]{fabrizio.carbone@epfl.ch}
\affiliation{Institute of Physics, Laboratory for Ultrafast Microscopy and Electron Scattering (LUMES), \'Ecole Polytechnique F\'ed\'eral de Lausanne, Station 6, CH-1015 Lausanne, Switzerland}

\begin{abstract}
Light-electron interaction in empty space is the seminal ingredient for free-electron lasers and also for controlling electron beams to dynamically investigate materials and molecules. Pushing the coherent control of free electrons by light to unexplored timescales, below the attosecond, would enable unprecedented applications in light-assisted electron quantum circuits and diagnostics at extremely small timescales, such as those governing intramolecular electronic motion and nuclear phenomena. We experimentally demonstrate attosecond coherent manipulation of the electron wave function in a transmission electron microscope, and show that it can be pushed down to the zeptosecond regime with existing technology. We make a relativistic pulsed electron beam interact in free space with an appropriately synthesized semi-infinite light field generated by two femtosecond laser pulses reflected at the surface of a mirror and delayed by fractions of the optical cycle. The amplitude and phase of the resulting coherent oscillations of the electron states in energy-momentum space are mapped via momentum-resolved ultrafast electron energy-loss spectroscopy. The experimental results are in full agreement with our theoretical framework for light-electron interaction, which predicts access to the zeptosecond timescale by combining semi-infinite X-ray fields with free electrons.
\end{abstract}
\date{\today}
% {\bf KEYWORDS:} xxx
%\pacs{xxx}
\maketitle

\section{Introduction}

The scattering of single photons by free electrons is extremely weak, as quantified by the Thomson scattering cross-section, which for visible frequencies is of the order of $10^{-29}$\, m$^2$. Additionally, direct photon absorption or emission by a free-space electron is forbidden due to energy-momentum mismatch. To circumvent these limitations and increase the probability of electron-photon interaction, a variety of methods have been devised\cite{paper149}. For example, the Kapitza-Dirac effect involves a conceptually simple configuration in which an electron intersects a light grating produced by two counter-propagating light beams of the same frequency\cite{FAB01,B07}. The interaction is then elastic and requires the electron to undergo an equal number of virtual photon absorption/stimulated-emission processes. When the absorbed and emitted photons differ in energy, the interaction results in frequency up- or down-conversion\cite{WKI16,KES17}, which is the basis of undulator radiation and free-electron lasers\cite{FGK1988,MT10}.

A direct single-photon emission/absorption process can also bridge the energy-momentum mismatch if either the electrons are not free (e.g., in photoemission from atoms/molecules\cite{DHK02} and solid surfaces\cite{MLA06}) or when a scattering structure generates evanescent light fields\cite{BFZ09} in the vicinity of the interaction volume. Such an electron-photon-matter interaction creates optical field components with a frequency-momentum decomposition that lies outside the light cone, allowing emission/absorption to take place. This type of interaction, which is forbidden in free space\cite{L1979,WHC17}, is regularly exploited for generating radiation and for accelerating charged particles. Recently, it has also prompted the development of photon-induced near-field electron microscopy (PINEM)\cite{BFZ09,BZ15,paper151,PLZ10}. In PINEM, an energetic electron beam interacts with the evanescent near fields surrounding an illuminated material structure. The interaction is particularly strong when the structure supports surface-plasmon polaritons (SPP) that are excited by short light pulses\cite{PLQ15,paper272}. Optical near-fields then produce coherent splitting of the electron wave function in energy space, giving rise to Rabi oscillations among electron quantum states separated by multiples of the photon energy\cite{FES15}. The microscopic details of the process are encoded in the electron wave function, which can be revealed via ultrafast electron energy-loss spectroscopy (EELS) and controlled using suitable illumination schemes\cite{EFS16,PRY17}.

\begin{figure}[H]
\centering
\includegraphics[width=1.0\textwidth]{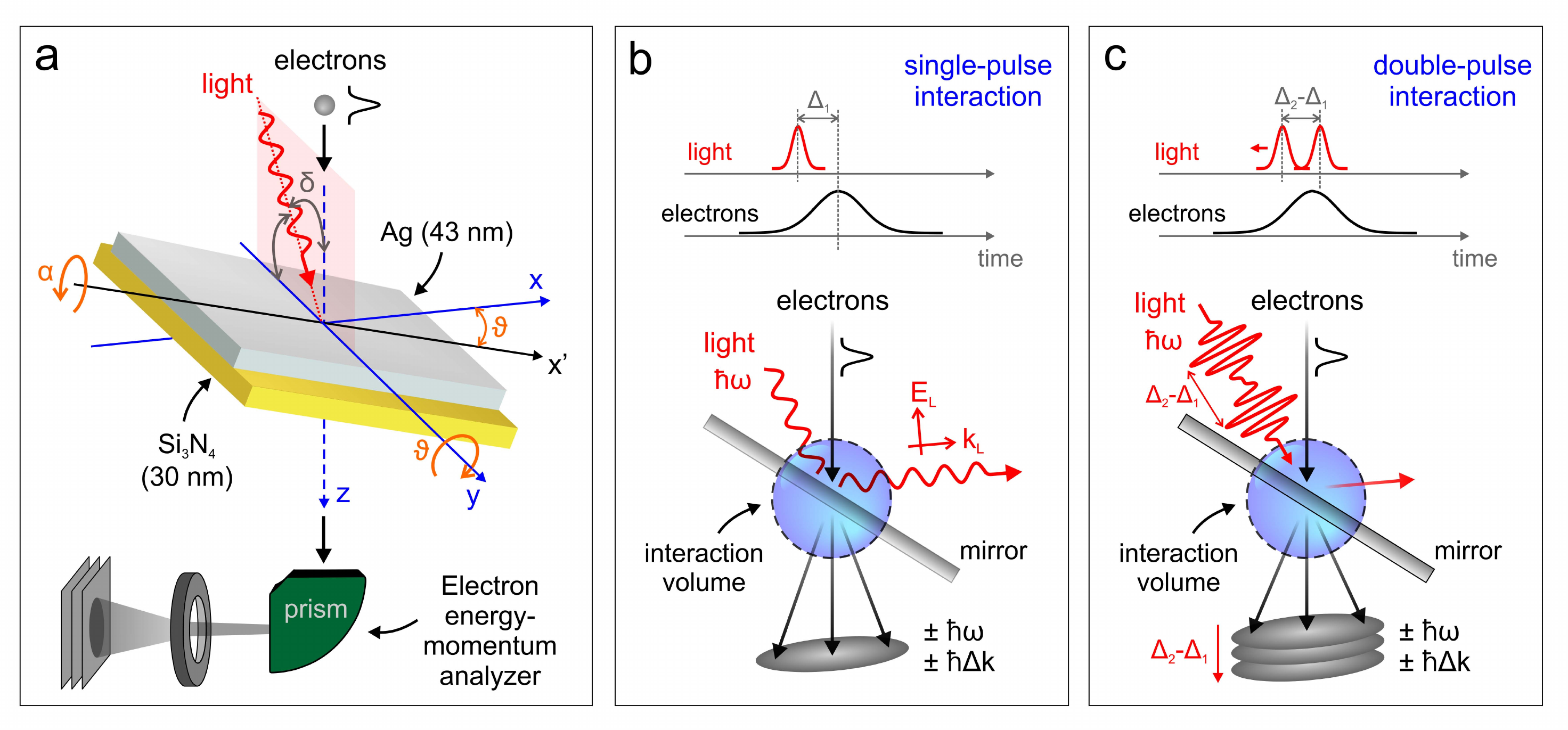}
\caption{{\bf Schematic representation of the experiment used to probe the interaction of free-electrons with semi-infinite light fields.} (a) Ultrashort 200\,keV electron pulses travel along the $z$ axis and impinge on the surface of a Ag/Si$_3$N$_4$ thin bilayer, which is mounted on a double-tilt holder able to rotate around the $x$ (angle $\alpha$) and $y$ (tilt angle $\vartheta$) axes. Light propagates within the $y$-$z$ plane, incident with an angle $\delta\sim4-5^\circ$ relative to the $z$ axis and then reflected from the Ag surface. The resulting electron-photon interaction is probed by monitoring electron energy-loss spectra as a function of geometrical parameters and light properties. (b) Description of the electron-light interaction here explored. The breaking of translational invariance produced by light reflection enables photon absorption or emission by the electron corresponding to a quantized energy and momentum exchange. (c) Description of the three-pulse experiment used for coherent modulation of the electron wave function. Electrons interact with an appropriately synthesized optical field distribution produced by two mutually-phase-locked photon pulses whose relative phase is changed by varying their relative delay $\Delta_2-\Delta_1$.}
\label{Fig1}
\end{figure}

In this work, we report on a different and more general method for controlling and manipulating the strength of electron-photon interaction. Instead of relying on localized near fields (e.g., plasmons), which inevitably depend on the intrinsic cross-section associated with the optical excitation of confined optical modes, we make use of a spatially abrupt interruption of the light field in free-space, also referred to as semi-infinite field\cite{PP1978,EP1979,AFM1979,PBC05,KGK14} (see Fig.\ \ref{Fig1}a). Such a boundary condition can be attained by sending the electrons through a light beam that intersects a refractor, an absorber, or more efficiently, a reflecting mirror along the optical path. When the light wave extends only over half-space, the energy-momentum conservation constraint is relaxed and electron-photon interaction can take place (see Fig.\ \ref{Fig1}b and Fig.\ S1 in Supporting Information (SI)) with an efficiency exceeding that produced by a resonant plasmonic nanostructure.

Within this scenario, we demonstrate attosecond coherent control of the electron wave function by appropriately synthesizing a semi-infinite optical field using a sequence of two mutually-phase-locked light pulses impinging on a mirror and delayed in time by fractions of the optical cycle (see schematics in Fig.\ \ref{Fig1}c). The profile of the field resulting from such a temporal combination of pulses changes the energy and momentum of an electron as it traverses the interaction volume. The energy-momentum distribution of electron states is recorded as a function of the delay between the two photon pulses \emph{via} momentum-resolved fs-EELS performed in an ultrafast transmission electron microscope\cite{PML15,VFZ16,FBR17}, revealing the light-induced modulation of both amplitude and phase of the electron wave function. Our experimental results are successfully described within a general theoretical framework for electron-light interaction, which is able to further predict the ability of this method to achieve coherent control over the electron wave function down to the zeptosecond regime using semi-infinite x-ray fields.

\section{Results and Discussion}

\begin{figure}[H]
\includegraphics[width=1.0\textwidth]{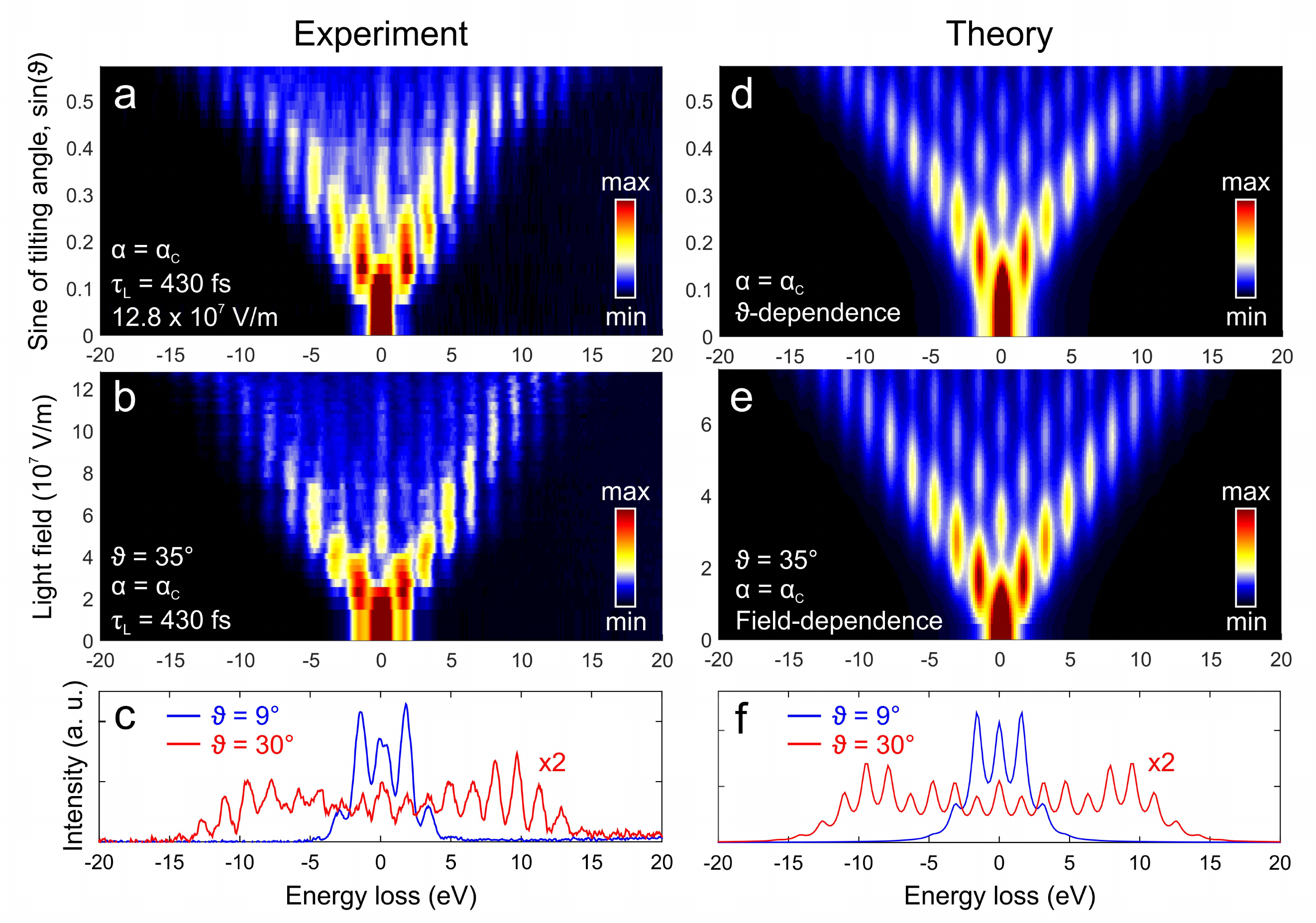}
\caption{{\bf Energy exchange during electron-light interaction.} (a) Sequence of measured EELS spectra (color map) plotted as a function of increasing angle $\vartheta$. We use p-polarized light (incident field along $x$ axis), $\alpha = \aC$, a peak field amplitude of $12.8\times10^7$\,V/m, and light and electron pulse durations $\tL = 430\,$fs and $\te = 410\,$fs. Sidebands at energies $\pm\ell\hbar \omega$ relative to the zero-loss peak (ZLP) are visible, where $\ell$ is the net number of exchanged photons. (b) Sequence of EELS spectra measured for increasing light field amplitude with fixed tilt angle $\vartheta=35^\circ$. (c) Spectra selected from (a), measured at $\vartheta = 9^\circ$ (blue curve) and $\vartheta = 30^\circ$ (red curve), showing a strong redistribution of the electron density toward the high-energy sidebands for large tilt angle. (d)-(f) Simulated EELS spectra corresponding to the experimental conditions of (a)-(c) (see Methods for details of calculations).}
\label{Fig2}
\end{figure}

The translational symmetry of a propagating electromagnetic wave is broken by refraction, absorption, or reflection at a material interface. In our study, we use a Ag thin film (43 nm) deposited on a Si$_3$N$_4$ membrane (30 nm) acting as a mirror. As schematically depicted in Fig.\ \ref{Fig1}a, the mirror is mounted on a double-tilt holder able to rotate around the $x$ (angle $\alpha$) and $y$ (tilting angle, $\vartheta$) axes. To demonstrate that electron-photon interaction can be strongly enhanced by the semi-infinite field effect, we display EELS spectra recorded as a function of laser field amplitude for a fixed orientation of the mirror (Fig.\ \ref{Fig2}b), and as a function of mirror tilting angle $\vartheta$ for fixed field amplitude (Fig.\ \ref{Fig2}a), using p-polarized light in all cases (incident field parallel to $x$ axis). Following the interaction, the zero-loss peak (ZLP) at an energy $E_0 = 200$\,keV is redistributed among sidebands at multiples of the incident photon energy $\pm\ell\hbar\omega$, corresponding to energy losses and gains by the electrons. At large values of both $\vartheta$ and the light field amplitude, the electron distribution is almost completely transferred toward high-energy spectral sidebands ($|\ell|\gg1$), leaving a nearly depleted ZLP and revealing a high probability for multiphoton creation and annihilation.

The modulation of the EELS spectra is determined by the integral of the optical electric field amplitude $\EE_z(z)$ along the electron beam direction $z$. Following previous works\cite{paper151,paper272,PLZ10}, the strength of the electron-photon interaction can be quantified in terms of the parameter (see Methods)
\begin{align}
\beta=(e\gamma/\hbar\omega)\int dz\;\EE_z(z)\,\ee^{-\ii\omega z/v}.
\label{betatext}
\end{align}
In particular, the fraction of electrons transmitted in the $\ell^{\rm th}$ sideband is approximately given by the squared Bessel function
\begin{align}
P_\ell=J^2_\ell(2|\beta|).
\label{Iell}
\end{align}
The spectral distribution of the electron density can be thus changed either by tilting the mirror (Fig.\ \ref{Fig2}b) or by increasing the laser power (Fig.\ \ref{Fig2}c), producing quantitatively similar effects. Considering the large permittivity ($\approx-30+0.4\ii$) of silver at the employed photon energy ($\hbar\omega\approx$1.57\,eV) and the small optical skin depth ($\approx11\,$nm for $1/e$ decay in intensity) compared with the silver layer thickness, the mirror reflects $>98\%$ of the incident light. Thus, neglecting light penetration inside the material, the electric field along the electron path can be considered to be made of incident (i) and reflected (r) components as $\EE_z(z)=\left(\EE_z^{\rm i}\ee^{\ii k_z^{\rm i} z}+\EE_z^{\rm r}\ee^{-\ii k_z^{\rm r} z}\right)\theta(-z)$, where the step function $\theta(-z)$ limits light propagation to the upper part of the mirror and $k_z^{\rm i/r}$ is the projection of the incident/reflected light wave vector along $z$. Inserting this field into Eq.\ (\ref{betatext}), we find
\begin{align}
\beta\approx(\ii e\gamma/\hbar\omega)\left[\frac{\EE_z^{\rm i}}{\omega/v-k_z^{\rm i}}+\frac{\EE_z^{\rm r}}{\omega/v+k_z^{\rm r}}\right],
\label{betatext2}
\end{align}
which makes the interaction strength finite and explicitly dependent on the field amplitude and tilting geometry. We further present in the Methods section a detailed analytical theory extended to deal with arbitrary pulse durations, two light pulses, and real material mirrors, used for comparison with the experimental results in the figures that follow. Nonetheless, Eq.\ (\ref{betatext2}) provides a satisfactory level of description that allows us to understand the data in simple terms, specially when the mirror is considered to be perfect (see Figs.\ S2 and S3 in SI).

Because light and electron beams in our apparatus are not collinear, the interaction strength described by $\beta$ for p-polarized light vanishes only when the tilt angles are set to $\vartheta = 0^{\circ}$ and $\alpha=\aC=12.9^\circ$, in agreement with calculations based on the theory reported on Methods. This corresponds to the condition that the incident and reflected amplitudes almost completely cancel each other in Eq.\ (\ref{betatext2}), hence producing a negligible net effect (minimum $|\beta|$, see red curve in Fig.\ S4, SI). This result is also in agreement with the relation $\aC=\tan^{-1}\left[\sin\delta/(\cos\delta-v/c)\right]$ derived in the SI from Eq.\ (\ref{betatext2}) to yield $\beta=0$ assuming a perfect mirror (blue curve in Fig.\ S4, SI). Likewise, $\beta$ cancels when the polarization is changed from p to s, a result that is clearly observed in polarization-dependent measurements (see Fig.\ S5 in SI).

To extract quantitative information on the measurements presented in Fig.\ \ref{Fig2}a-c, we perform the corresponding simulations shown in Fig.\ \ref{Fig2}d-f for the energy distribution of a pulsed electron beam after impinging on an illuminated Ag/Si$_3$N$_4$ bilayer film, using the same layer thicknesses and geometrical arrangement as in the experiment. In particular, we consider p-polarized light incident with $\alpha$ fixed to the critical angle $\aC$. Simulations are carried out incorporating realistic dielectric data for the involved materials (see Methods). The ratio of electron-to-light pulse durations $\te/\tL\approx410\,{\rm fs}/430\,{\rm fs}\approx0.95$ is the same as estimated in experiment (see Methods), long enough to ensure large temporal overlap between the electron and light pulses, thus enhancing the probability of interaction. The agreement between experiment and theory is rather satisfactory. Similar conclusions are also obtained from measurements and simulations for small $\tL$ compared with $\te$ (see Figs.\ S3 and S6 in SI).

We remark that, in contrast to previous studies of electron-photon interactions\cite{BZ15}, the effect here observed is primarily due to electrons coupling directly to the light waves rather than to the near-field created around a nanostructure. The kinematic mismatch in the electron-light coupling is remedied by the formation of semi-infinite light plane-waves (see Fig.\ S1 in SI). As noted above, at a photon energy of $\approx1.57\,$eV the silver skin depth ($\approx11\,$nm) is much smaller than both the optical wavelength and the metal layer thickness, so the evanescent tail inside the Ag film gives a negligible contribution, as confirmed by direct comparison with perfect-mirror simulations based on Eq.\ (\ref{betatext2}) (see Fig.\ S3 in SI).

Overall, this experiment-theory framework is general and allows describing other interesting scenarios, such as the phase-controlled combination of two interactions arising from semi-infinite light fields and plasmon polaritons propagating on a metal film. This is illustrated by measurements presented in Fig.\ S9 (SI), with SPPs generated at the edge of a nanocavity carved in the Ag layer. The interference between the traveling plasmon wave and the semi-infinite light field creates a standing wave distribution sampled by the electrons, which allows us to produce a snapshot of the SPP in real-space.

\begin{figure}[H]
\includegraphics[width=1.0\textwidth]{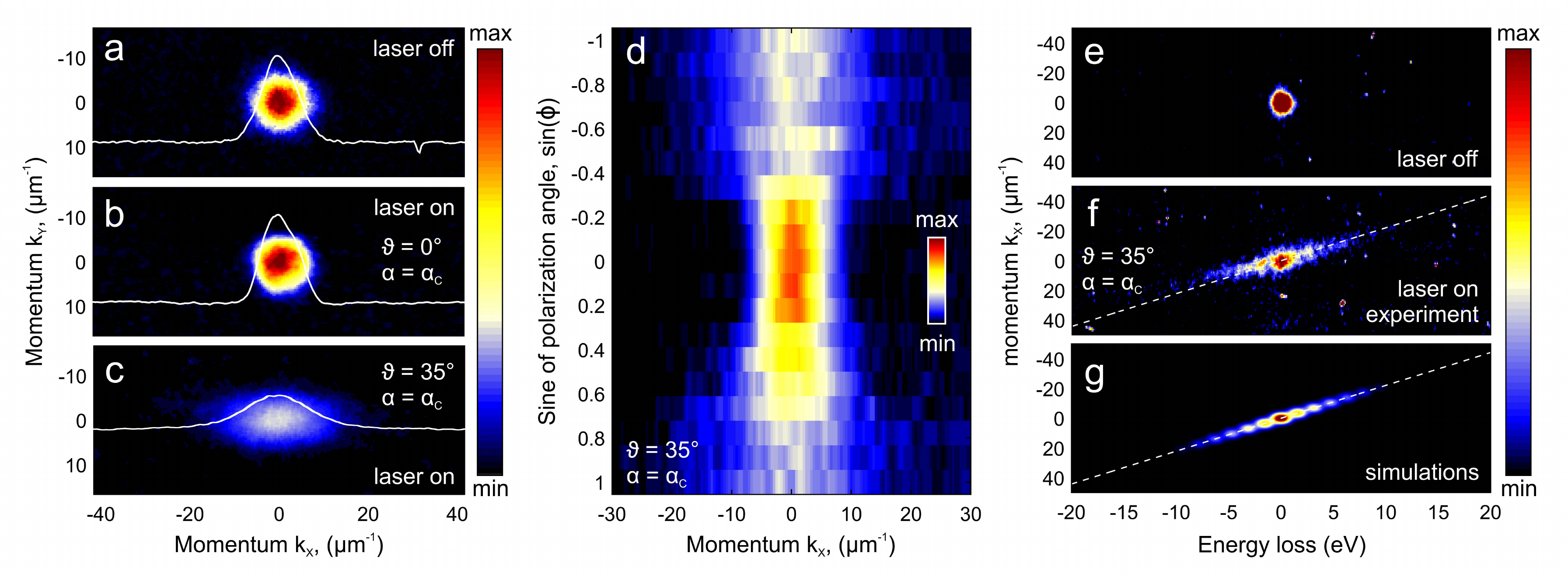}
\caption{{\bf Momentum exchange during electron-light interaction.} (a) Direct electron beam measured in the diffraction plane as a function of transversal momentum  $(k_x,k_y)$ when no light is applied. (b)-(c) Same as (a) under illumination with 560\,fs laser pulses of $11.1\times10^{7}$\,V/m peak field amplitude and $\alpha = \aC$. The $\vartheta$ tilt angle is $0^\circ$ in (b) and $35^\circ$ in (c). A clear streaking of the electron beam appears along the $k_{x}$ direction in (c) as a result of momentum exchanges between light and electrons. (d) Electron beam profile along $k_{x}$ as a function of light polarization ($\sin\phi=0$ for s-polarization and $\sin\phi=\pm1$ for p-polarization). (e) Direct electron beam measured in the momentum-energy plane $k_{x}$-$E$ in the absence of optical illumination. (f)-(g) Measured (f) and simulated (g) momentum-energy maps for illumination under the conditions of (c).}
\label{Fig3}
\end{figure}

Energy exchanges between light and electrons should be also accompanied by momentum transfers along the direction parallel to the film, where translational invariance guarantees momentum conservation. Measuring such momentum exchanges is quite challenging because of the small induced electron deflection (only a few ${\rm \mu}$rad), which demands high transverse coherence that we achieve by operating the microscope in high dispersion diffraction mode. In Fig.\ \ref{Fig3}a we show the direct electron beam measured in the $k_{x}$-$k_{y}$ diffraction plane when no light is applied, whereas Fig.\ \ref{Fig3}b,c shows the effect of light interaction for tilt angles $\vartheta = 0^\circ$ and $\vartheta = 35^\circ$, with fixed $\alpha = \aC$. A clear streaking of the electron beam appears along the $k_{x}$ direction for $\vartheta = 35^\circ$ as a result of the noted momentum exchange. As already observed in the electron energy spectra, the interaction vanishes at $\vartheta = 0$ and $\alpha = \aC$ for p-polarization, resulting in zero momentum exchange. The physical origin of this behavior is well described by the analytical expressions in Eqs.\ (\ref{Iell}) and (\ref{betatext2}), in which the electric field component along the z axis modulates the interaction strength. This can be also experimentally probed by rotating the polarization of the light wave, which results in a corresponding modulation of the electron beam streaking (see Fig.\ \ref{Fig3}d). Particularly instructive is the simultaneous visualization of inelastic energy and momentum exchanges, which we directly map using the reciprocal-space imaging ability of the electron spectrometer in our microscope\cite{PML15,PLQ15} (see Fig.\ \ref{Fig3}e-g). The streaking of the electron beam occurs along a line $\hbar\omega=\hbar c q_{{\rm T},x}$, where $q_{{\rm T},x}$ is the transverse component of the transferred momentum along $x$, which in the limit of small angles $\delta$ and $\alpha$ admits the expression $q_{{\rm T},x} \approx (\omega/c) \cos\vartheta\,\sin\vartheta$ (see SI for the full derivation). For every photon absorption/emission event the electron gains/loses a quantum of energy $\hbar\omega$ and momentum $\hbar q_{{\rm T},x}$ along $x$. The unique ability of our technique to map transient energy exchanges in momentum space could prompt the development of new microscopy methods in which the limitation imposed by EELS energy resolution is lifted for large momentum transfers, such as in the dynamic imaging of low-energy phonons.

These results provide a full characterization of electron-photon interaction at the mirror interface in energy-momentum space, which suggests using such interaction for the coherent manipulation of the electron wave function. We implement this idea by engineering the parameter $|\beta|$ (which can be thought of as a light-driven Rabi phase for transitions in the electron multilevel quantum ladder with $\hbar\omega$ energy spacings\cite{FES15}) through a three-pulse experiment in which the electron interacts with a properly shaped field distribution consisting of a sequence of two mutually-phase-locked photon pulses, delayed by time intervals $\Delta_1$ and $\Delta_2$ with respect to the electron pulse (see schematics in Fig.\ \ref{Fig1}c and additional details in Methods). We change the relative phase between the two light pulses by varying $\Delta_2-\Delta_1$ in steps of 500 attoseconds. The field distribution resulting from such a temporal combination of pulses is then used to coherently manipulate the energy-momentum distributions of the electrons.

\begin{figure}[H]
\includegraphics[width=0.95\textwidth]{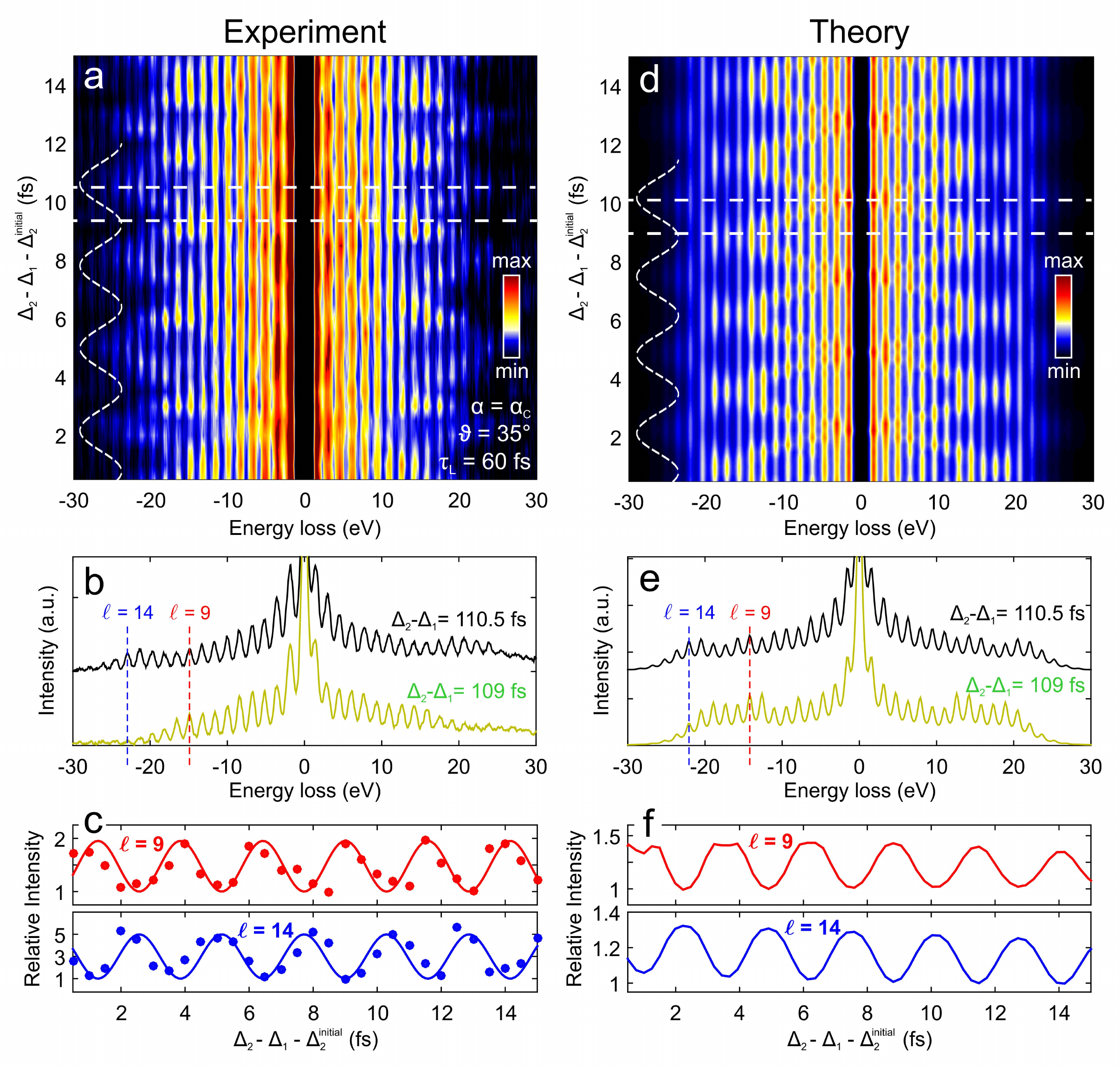}
\caption{{\bf Attosecond coherent control of free-electrons.} The electron beam interacts with a semi-infinite temporally modulated optical field distribution produced by a sequence of two mutually-phase-locked light pulses impinging on the mirror. (a) Measured EELS spectra as a function of relative delay $\Delta_2-\Delta_1$ between the two optical pulses. The tilt angles are $\vartheta=35^\circ$ and $\alpha=\aC$, the optical pulses are 60\,fs long with a peak field amplitude of $21.4\times10^{7}\,$V/m each, and the delays are $\Delta_1=0$ and $\Delta_2\approx100-115$\,fs, with $\Delta^{\rm initial}_2 = 100$\,fs. (b) EELS spectra taken at two different time delays (marked by horizontal dashed lines in panel (a)). (c) Relative intensity (full circles) of the $\ell=9$ and $\ell=14$ sidebands plotted as a function of time delay between the two optical pulses, exhibiting a periodic modulation of period $\approx2.6\,$fs (equal to the optical cycle $2\pi/\omega$) and a relative $\pi$ phase shift. Solid curves are least-square fits to the data. (d)-(f) Simulated EELS spectra and resulting intensity change corresponding to the experimental conditions of (a)-(c) (see Methods for details of calculations).}
\label{Fig4}
\end{figure}

A sequence of EELS spectra measured as a function of $\Delta_2-\Delta_1$ is shown in Fig.\ \ref{Fig4}a for $\vartheta=35^\circ$, $\alpha=\aC$, $\te\approx350\,$fs electron pulses, $\tL\approx60$\,fs optical pulses, a light field amplitude of $21.4\times10^{7}\,$V/m per pulse, and delays $\Delta_1 = 0\,$fs and $\Delta_2\approx100-115\,$fs. The large values of $\Delta_2$ enable fine modulation of the optical phase while considerably reducing the intensity changes associated with light-pulse overlap. We observe periodic oscillations of the spectral sidebands with a period $\approx2.6\,$fs equal to the optical cycle $2\pi/\omega$. This effect cannot be assimilated to a simple intensity variation of the impinging light, which stays at the $\sim 5\times10^{-2}$ level; in fact, the employed intensities lie in the saturation regime, as shown by the negligible electron-spectra changes observed for single light pulses at high field amplitude (see Fig.\ S7 in SI). Detailed inspection of the EELS spectra for two different delays ($\Delta_2 = 109\,$fs and 110.5\,fs in Fig.\ \ref{Fig4}b, corresponding to the horizontal dashed lines in Fig.\ \ref{Fig4}a) reveals radically different distributions of the sidebands relative to the ZLP, which are further quantified in Fig.\ \ref{Fig4}c by plotting the $\ell=9$ and $\ell=14$ features as a function of $\Delta_2-\Delta_1$. We observe significant intensity oscillations with a period of $\approx2.6\,$fs and a well-defined $\sim\pi$ relative phase shift. We remark once more that measurements shown in Figs.\ \ref{Fig4}a and \ref{Fig4}b,c are well reproduced by our analytical simulations for two light pulses (see Methods) plotted in Figs.\ \ref{Fig4}d and \ref{Fig4}e,f, respectively.

\begin{figure}[H]
\includegraphics[width=1.0\textwidth]{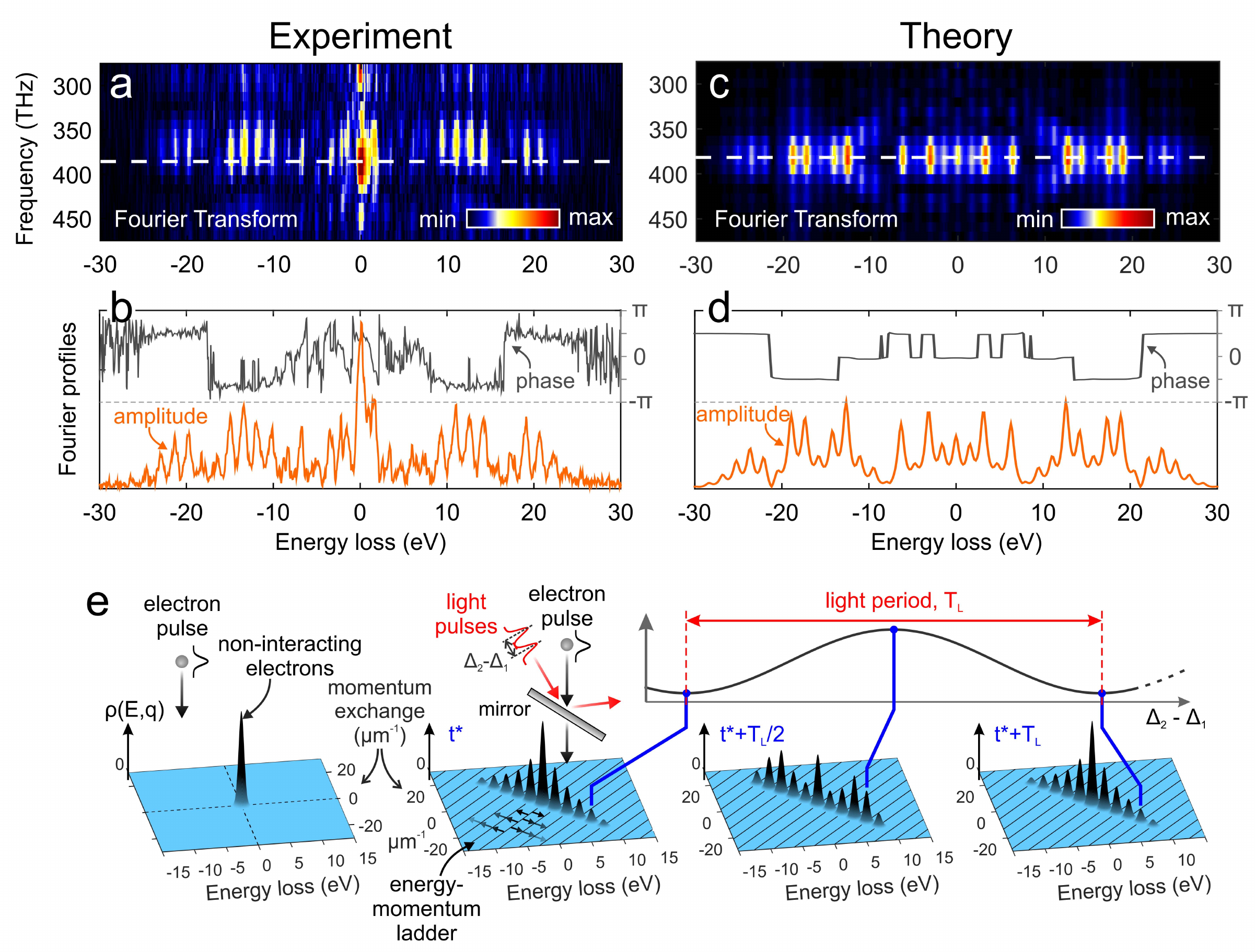}
\caption{{\bf Amplitude and phase modulation of the electron wave function.} (a) Two-dimensional Fourier transform of the energy-time map plotted in Fig.\ \ref{Fig4}a. (b) Complex spectral distribution of the electron-wave-function amplitude (bottom) and phase (top), extracted at the modulation frequency of $2\pi/(2.6\,{\rm fs})\approx385\,$THz. (c)-(d) Two-dimensional Fourier transform extracted from the calculated energy-time map plotted in Fig.\ \ref{Fig4}d. (e) Schematic representation of electron-wave-function modulation, showing snapshots of the strong energy-momentum electron density redistribution for different values of the phase shift between the two optical pulses.}
\label{Fig5}
\end{figure}

This behavior is indicative of a continuous redistribution within the quantum electron-population ladder, periodically transferred back and forth between high- and low-energy levels. Such an effect is the result of coherent modulation of the electron wave function \emph{via} the coherent constructive and destructive modulation of $|\beta|$ when changing the relative phase between the two driving optical pulses. The time-Fourier transform of the maps in Figs.\ \ref{Fig4}a and \ref{Fig4}d, presented in Figs.\ \ref{Fig5}a and \ref{Fig5}c, gives access to the spectral distribution within the quantum ladder at the modulation frequency $2\pi/(2.6\,{\rm fs})\approx385\,$THz. The amplitude and phase of such a modulation, shown in Figs.\ \ref{Fig5}b and \ref{Fig5}d, provide a complete picture of the optically-manipulated electron wave function resolved for each electron energy level.

The coherent control of ultrafast electron beams has recently attracted much attention for its potential application in novel ultrashort (attosecond) electron sources, as well as electron imaging and spectroscopy. While semi-infinite light beams have been used for the temporal streaking and compression of electron pulses\cite{KSE16,RB16,MB17}, their potential to control purely quantum aspects of the electron energy-momentum distribution has not been fully explored. This approach allows us to develop new capabilities of coherent control of free electrons beyond the utilization of surface and localized plasmons employed so far to assist the electron scattering\cite{EFS16,PRY17}. In our experiments, we synthesize a semi-infinite temporally modulated field distribution (obtained by a sequence of two mutually-phase-locked light pulses impinging on a mirror) to demonstrate coherent modulation of the electron wave function. A schematic representation of such modulation is shown in Fig.\ \ref{Fig5}e, where snapshots of the strong electron density redistribution in both energy and momentum, as observed experimentally and calculated theoretically, are presented for different values of the optical phase shift of the synthesized optical field distribution.

This method, which does not involve evanescent near fields, is very general and only requires a refracting, absorbing, or reflecting interface. The experimental requirements are thus simplified and do not necessitate nanofabricated structures for the excitation of plasma resonances, while potentially enabling electro-optical tunability through graphene gating.

\begin{figure}[H]
\includegraphics[width=1.0\textwidth]{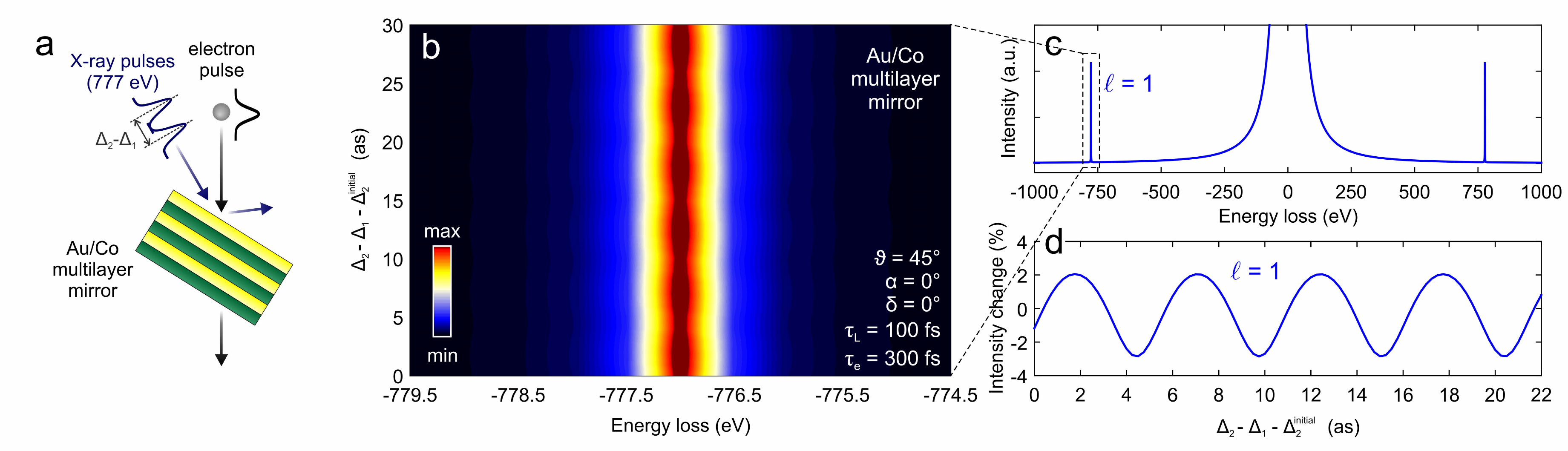}
\caption{{\bf Toward zeptosecond coherent control of free electrons.} (a) An electron beam interacts with a semi-infinite temporally modulated x-ray field ($777\,$eV photon energy) produced by a sequence of two mutually-phase-locked pulses partially reflected by a Au/Co multilayer. (b) Calculated EELS spectra as a function of relative delay $\Delta_2-\Delta_1$ between the two x-ray pulses. The tilt angles are set to $\vartheta=45^\circ$ and $\alpha=0^\circ$, while the x-ray pulses are 100\,fs long with a peak field amplitude of $9.4\times10^{9}\,$V/m per pulse. Simulations are performed within a 30-attosecond window starting from $\Delta^{\rm initial}_2 = 150$\,fs ($\Delta_1=0$). (c) Calculated EELS spectrum showing the electron sidebands at energies of $\pm777\,$eV with respect to the ZLP. (d) Relative intensity change of the $\ell=1$ sideband plotted as a function of time delay between the two x-ray pulses, exhibiting a periodic modulation of period $\approx5.3\,$as (equal to the x-ray cycle) and an intensity change rate of $\approx1\,\%$ per $511\,zs$.}
\label{Fig6}
\end{figure}

A particularly appealing possibility consists in controlling electron-light interactions using photons of different energies, not restricted by the ability of materials to support localized resonances such as plasmon polaritons, but solely determined by the quality of the mirror surface at a specific frequency. Using high-energy photons all the way to the x-ray regime, our methodology would then allow us to control the electron wave function down to the zeptosecond timescale\cite{HPP13}. To verify the feasibility of this idea, we have designed a multilayer mirror composed of 30 layers of 1.6-nm-thick cobalt spaced by 1-nm-thick gold (total thickness is 78\,nm), still transparent for 200\,KeV electrons and capable of reflecting around $35\%$ of 777\,eV light at an angle of incidence of $45^\circ$ (see Fig.\ S10 in the SI). This type of mirror is routinely used in x-ray facilities\cite{KD1989}. We then simulated a three-pulse experiment with two 100\,fs, 777\,eV, 50\,TW/cm$^2$ x-ray pulses, similar to what is currently available from free-electron lasers\cite{DESY2017}, impinging on the multilayer along the same direction as a 300\,fs electron pulse (see schematics in Fig.\ \ref{Fig6}a). We carry out simulations within a 30-attosecond window starting from an initial delay $\Delta_2-\Delta_1 = 150\,$fs. Electron sidebands are clearly discernable at energies of $\pm777\,$eV relative to the ZLP (see Fig.\ \ref{Fig6}c), originating in the same electron-ladder interaction as observed for near-infrared light. The resulting EELS spectrum as a function of the delay between the two x-ray pulses is displayed in Fig.\ \ref{Fig6}b, while the relative intensity change for the first sideband is shown in Fig.\ \ref{Fig6}d, revealing a clear modulation by the optical cycle of the x-ray pulse ($\approx5.3\,$as) and an intensity change rate of $\approx1\,\%$ per $511\,zs$. Coherent manipulation of the electron wave function can be thus pushed to the zeptosecond regime using currently existing technology within our electron-light interaction scheme. Access to such timescales may open new perspectives for the observation of intramolecular electronic motions\cite{OSS17} and nuclear processes such as fission, quasifission, and fusion\cite{DHD11}.

\newpage

\section*{METHODS}

{\bf Materials and Experiment.} A sketch of our experiment is depicted in Fig.\ \ref{Fig1}a. We used an ultrafast transmission electron microscope (a detailed description can be found in Ref.\ \onlinecite{PML15}) to focus femtosecond electrons and light pulsed beams on an optically-thick mirror. The mirror was thin enough to transmit the electrons while producing large light reflection. Specifically, it was made of a 43\,nm-thick ($\pm5\,$nm) silver thin film sputtered on a 30\,nm Si$_3$N$_4$ membrane placed on a Si support with a $80\times80\,\mu$m$^2$ window, which was in turn mounted on a double-tilt sample holder that ensured rotation around the $x$ (angle $\alpha$) and $y$ (angle $\vartheta$) axes over a $\pm35^\circ$ range. Electron pulses were generated by photoemission from a UV-irradiated LaB$_6$ cathode, accelerated to an energy $E_0=200\,$keV along the $z$ axis, and focused on the specimen surface. The mirror was simultaneously illuminated with femtosecond laser pulses of $\hbar\omega=1.57\,$eV central energy and variable duration, intensity, and polarization. The light pulses were focused on the sample surface (spot size of $\sim58\,\mu$m FWHM). The light propagation direction lied within the $y$-$z$ plane and formed an angle $\delta\sim4-5^\circ$ with the $z$ axis, as shown in Fig.\ \ref{Fig1}a. The delay between electrons and photons was varied via a computer-controlled delay line. For the three-pulse experiment, we implemented a Michelson interferometer along the optical path of the infrared beam, incorporating a computer-controlled variable delay stage on one arm.

The transmission electron microscope was equipped with EELS capabilities, coupled to real-space and reciprocal-space imaging. Energy-resolved spectra were acquired using a Gatan imaging filter (GIF) camera operated with a 0.05\,eV-per-channel dispersion setting and typical exposure times of the CCD sensor from 30\,s to 60\,s. Multiple photon absorption and emission events experienced by the electrons were analyzed as a function of relative beam-mirror orientations by recording EELS spectra and diffraction patterns in high-dispersion-diffraction mode. During post-acquisition analysis, the EELS spectra were aligned based on their ZLP positions using a differential-based maximum intensity alignment algorithm.

Special care was taken in evaluating the temporal width of the light and electron pulses. An infrared auto-correlator was used for measuring the duration of the infrared pulses. For electrons, the pulse duration was estimated by measuring the electron-photon cross-correlation as obtained by monitoring the EELS spectra as a function of the delay time between electrons and the infrared light. In the low-excitation regime, the measured temporal width of the $\ell^{\rm th}$ sideband is roughly $\tau_\ell\approx\sqrt{(\te^2+(\tL^2)/\ell)}$ (i.e., the convolution of electron and optical pulses\cite{PPZ14} of durations $\te$ and $\tL$, respectively). For infrared pulses with $\tL=\,$60\,fs, 175\,fs, and 430\,fs FWHM, we derived electron pulse durations $\te=\,$350\,fs, 395\,fs, and 410\,fs FWHM, respectively ($<5\%$ estimated error).

{\bf Theory of ultrafast electron-light interaction.} Following previous works\cite{paper151,paper272,PLZ10}, we describe an electron wave-packet exposed to an optical field through the Schr\"odinger equation $(H_0+H_1)\psi=\ii\hbar\partial\psi/\partial t$, where $\psi(\rb,t)$ is the electron wave function, $H_0$ is the free-space Hamiltonian, and $H_1=(-\ii e\hbar/\me c)\Ab(\rb,t)\cdot\nabla$ represents the minimal-coupling interaction involving the optical vector potential $\Ab(\rb,t)$ in a gauge in which the scalar potential and $\nabla\cdot\Ab$ are both zero. We consider an expansion of the electron wave function in terms of components $\ee^{\ii(\kb\cdot\rb-E_\kb t/\hbar)}$ of momentum $\hbar\kb$ piled near a central value $\hbar\kb_0$ with $k_0=\hbar^{-1}\sqrt{(2\me E_0)(1+E_0/2\me c^2)}$, corresponding to an electron kinetic energy $E_0$. Each of these components is an eigenstate of $H_0$ with energy $E_\kb\approx E_0+\hbar\vb\cdot(\kb-\kb_0)$, where $\vb=(\hbar\kb_0/\me)/(1+E_0/\me c^2)$ is the central electron velocity. This approximation is valid for small momentum spread (i.e., $|\kb-\kb_0|\ll k_0$). Under these conditions, we can also approximate $H_0\approx E_0-\hbar\vb\cdot(\ii\nabla+\kb_0)$, as well as $\nabla\approx\ii\kb_0$ in $H_1$. Now, it is convenient to separate the fast evolution of the wave function imposed by the central-momentum component as $\psi(\rb,t)=\ee^{\ii(\kb_0\cdot\rb-E_0t/\hbar)}\phi(\rb,t)$, where $\phi(\rb,t)$ then displays a slower dynamics. Putting these elements together, the Schr\"odinger equation reduces to
\begin{align}
(\vb\cdot\nabla+\partial/\partial t)\,\phi=\frac{-\ii e\gamma\vb}{\hbar c}\cdot\Ab\,\phi,
\nonumber
\end{align}
where $\gamma=1/\sqrt{1-v^2/c^2}$,
which admits the rigorous solution
\begin{widetext}
\begin{align}
\phi(\rb,t)=\phi_0(\rb-\vb t)\,\exp\left[\frac{-\ii e\gamma\vb}{\hbar c}\cdot\int_{-\infty}^t dt'\,\Ab(\rb+\vb t'-\vb t,t')\right].
\label{solution0}
\end{align}
\end{widetext}
Here, $\phi_0(\rb-\vb t)$ is the electron wave function before interaction with the optical field. In practice, we consider illumination by an optical pulse with a narrow spectral distribution centered around a frequency $\omega$, so the vector potential can be approximated as $\Ab(\rb,t)\approx(-\ii c/\omega)\EEb_0(\rb,t)\ee^{-\ii\omega t}+{\rm c.c.}$, where the electric field amplitude $\EEb_0(\rb,t)$ describes a slowly-varying pulse envelope that changes negligibly over an optical period. Inserting this expression into Eq.\ (\ref{solution0}), we find the solution
$\phi(\rb,t)=\phi_0(\rb-\vb t)\,\ee^{-\mathcal{B}+\mathcal{B}^*}$, where $\mathcal{B}(\rb,t)=\frac{e\gamma\vb}{\hbar\omega}\cdot\int_{-\infty}^t dt'\,\EEb_0(\rb+\vb t'-\vb t,t')\,\ee^{-\ii\omega t'}$. Finally, using the Jacobi-Anger expansion $\ee^{\ii u\sin\varphi}=\sum_{\ell=-\infty}^\infty J_\ell(u)\ee^{\ii\ell\varphi}$ (see Eq.\ (9.1.41) of Ref.\ \onlinecite{AS1972}) with $|u|=2|\mathcal{B}|$ and $\varphi=\arg\{-\mathcal{B}\}$, we obtain $\phi(\rb,t)=\phi_0(\rb-\vb t)\sum_{\ell=-\infty}^\infty J_\ell(2|\mathcal{B}|)\,\ee^{\ii\ell\arg\{-\mathcal{B}\}}$. This expression has general applicability under the assumptions of small energy spread in both electron and optical pulses.

For monochromatic light (i.e., when $\EEb_0(\rb)$ depends only on position), considering without loss of generality $\vb$ along $\zz$, we find $\mathcal{B}=\beta(\rb)\ee^{-\ii\omega(z/v-t)}$ with
\begin{align}
\beta(\rb)=\frac{e\gamma}{\hbar\omega}\int_{-\infty}^{z} dz'\,\EE_{0z}(x,y,z')\,\ee^{-\ii\omega z'/v},
\label{beta}
\end{align}
and the electron wave function then becomes
\begin{align}
\phi(\rb,t)=\phi_0(\rb-\vb t)\sum_{\ell=-\infty}^\infty J_\ell(2|\beta|)\,\ee^{\ii\ell\arg\{-\beta\}+\ii\ell\omega(z/v-t)},
\label{solution1}
\end{align}
where the last term in the exponential shows a change in the energy and momentum of the $\ell$ wave-function component given by $\ell\hbar\omega$ and $\ell\hbar\omega/v$.

For a Gaussian light pulse $\EEb_0(\rb,t)=\EEb_0(\rb)\ee^{-t^2/\sL^2}$, corresponding to a FWHM-intensity duration $\tL = (\sqrt{2\log2})\sL\approx1.18\sL$, under the assumption that the time needed by the electron to cross the interaction region is small compared with $\sL$, we recover the result of Eq.\ (\ref{solution1}) with $\beta$ (Eq.\ (\ref{beta})) replaced by $\ee^{-(z/v-t)^2/\sL^2}\beta$.

We now calculate the electron probability at the detector as the integral $\int d^3\rb|\phi(\rb,t)|^2$ for a large time $t$. Assuming a Gaussian electron pulse $\phi_0(\rb-\vb t)\propto\ee^{-(t-z/v-\Delta_1)^2/\se^2}$ normalized to one electron ($\int d^3\rb|\phi_0|^2=1$), with FWHM-intensity duration $\te = (\sqrt{2\log2})\se$, and a delay $\Delta_1$ relative to the light pulse, we find the probability that the electron has exchanged a net number of photons $\ell$ to be
\begin{align}
P_\ell=\sqrt{\frac{2}{\pi}}\frac{1}{\se}\int dt\;\ee^{-2t^2/\se^2}J_\ell^2\left(2|\beta|\ee^{-(t+\Delta_1)^2/\sL^2}\right),
\label{Pl}
\end{align}
with $\beta$ evaluated in the $z\rightarrow\infty$ limit of Eq.\ (\ref{beta}). From the identity\cite{AS1972} $\sum_\ell J_\ell^2(u)=1$, we reassuringly obtain $\sum_\ell P_\ell=1$. In the derivation of this expression, we have assumed that different $\ell$ electron channels have well separated energies, a condition that is guaranteed by the assumption of small energy spread in both pulses (i.e., $E_0\se\gg\hbar$ and $\omega\sL\gg1$). Finally, using the Taylor expansion $J_\ell(u)=\sum_{j=0}^{\infty}(-1)^j(u/2)^{\ell+2j}/j!(\ell+j)!$ for the Bessel functions\cite{AS1972}, the time integral in Eq.\ (\ref{Pl}) can be readily performed term by term to yield
\begin{align}
P_\ell=\sum_{j=0}^\infty\sum_{j'=0}^\infty C_{\ell j}C_{\ell j'}
\;\frac{1}{\sqrt{\lambda}}\;\ee^{-2n(\Delta_1^2/\sL^2)/\lambda},
\label{Plone}
\end{align}
where $n=\ell+j+j'$, $\lambda=1+n(\se/\sL)^2$, and $C_{\ell j}=(-1)^j|\beta|^{\ell+2j}/j!(\ell+j)!$ In the monochromatic limit $\sL\gg\se$, we trivially obtain $P_\ell=J_\ell^2(2|\beta|)$ (i.e., Eq.\ (\ref{Iell})).

Under illumination with two identical light pulses delayed by $\Delta_i$ ($i=1,2$) relative to the electron and with their amplitudes scaled by real factors $A_i$, a similar analysis can be carried out, using the Newton binomial expansion, to yield
\begin{align}
P_\ell=\sum_{j=0}^\infty\sum_{j'=0}^\infty\sum_{s=0}^N\sum_{s'=0}^N
&C_{\ell j}C_{\ell j'}
\binom{n}{s}\binom{n}{s'}
A_1^{2n-s-s'}A_2^{s+s'}\cos\left[(s-s')\omega(\Delta_2-\Delta_1)\right] \label{Pltwo}\\
&\times
\frac{1}{\sqrt{\lambda}}\;\ee^{-2n(\Delta_{12}^2/\sL^2)/\lambda}
\;\ee^{-[1-2(s-s')/2n](s-s')n(\Delta_2-\Delta_1)^2/\sL^2)},
\nonumber
\end{align}
where $\Delta_{12}=\left[(2n-s-s')\Delta_1+(s+s')\Delta_2\right]/2n$.

In our numerical simulations, we use Eqs.\ (\ref{Plone}) and (\ref{Pltwo}) with the electric field obtained by a standard transfer-matrix approach for a bilayer formed by Ag and Si$_3$N$_4$, with the permittivities of these materials taken as\cite{JC1972} $-30.3+0.39\ii$ and\cite{P1985} $4.04$, respectively. Calculations for x-ray pulses at 777\,eV photon energy are performed for multilayers of Au and Co, described by their permittivities $0.97+0.014\ii$ and $1.01+ 0.0014\ii$, respectively. Light amplitudes in the simulations are reduced by a factor of 1.7 with respect to the experimental estimates. This factor, which provides the best theory-experiment fit, is presumably originating in unaccounted losses along the optical path of the laser beam (see SI).

\section*{Acknowledgements} 
This work has been supported in part by the NCCR MUST of the Swiss National Science Foundation and the Spanish MINECO (MAT2014-59096-P and SEV2015-0522), AGAUR (2014 SGR 1400), Fundaci\'o Privada Cellex, and the Catalan CERCA program.The authors acknowledge Dr P. Baum and Dr C. Ropers for useful discussions. 

\newpage
\section*{FIGURE CAPTIONS}

%\begin{widetext} \begin{sidewaysfigure}
 %\end{sidewaysfigure} \end{widetext}

\newpage %\begin{widetext} \begin{sidewaysfigure}
% \end{sidewaysfigure} \end{widetext}

\newpage %\begin{widetext} \begin{sidewaysfigure}
 % \end{sidewaysfigure} \end{widetext}

\newpage % \begin{widetext} \begin{sidewaysfigure}
 % \end{sidewaysfigure} \end{widetext}

\newpage %\begin{widetext} \begin{sidewaysfigure}
 %\end{sidewaysfigure} \end{widetext}

\newpage % \begin{widetext} \begin{sidewaysfigure}
 % \end{sidewaysfigure} \end{widetext}

%\bibliographystyle{apsrev}
%\bibliography{../../bibtex/refs}

\end{document}